\documentclass[epj]{svjour}
\usepackage{graphics}
\usepackage{graphicx}
\usepackage{indentfirst}
\usepackage{color}
\usepackage{epsfig}
\begin{document}

\title{Investigation of induced fission of $^{nat}$Pb 
by accelerated $^{7}$Li ions}

\author{N. A. Demekhina \inst{1} \and G. S. Karapetyan\inst{2} \and
V. Guimar\~aes\inst{2}}

\institute{Yerevan Physics Institute, Alikhanyan Brothers 2, Yerevan 0036, 
Armenia\\
Joint Institute for Nuclear Research (JINR), Flerov Laboratory of Nuclear 
Reactions (LNR), Joliot-Curie 6, Dubna 141980, Moscow region Russia 
\and Instituto de Fisica, Universidade de S\~ao Paulo, P. O. Box 66318, 
05389-970 S\~ao Paulo, SP, Brazil}

\abstract{
Cross-sections for fragments produced by the
fission reaction induced by accelerated $^{7}$Li ions impinged
into an $^{nat}$Pb target at 245 MeV were obtained.
We have applied the  induced-activation method in off-line analysis. 
The analysis of charge and mass distributions of fission products 
allowed the determination of the total fission cross-section
for the $^7$Li+Pb system. 
The recoil technique (``thick target- thick catcher''), based on the two 
step model mathematical formalism, is used to determine the kinematics 
characteristics of reaction products.  Analysis concerning transferred 
linear momentum provided information on the initial projectile-target 
interaction, and is compared with proton-induced fission measurements. 
}


\maketitle

\section{Introduction}

The investigation of nucleon-nucleus and nucleus-nucleus collision mechanisms 
at low and intermediate energies is a source of understanding the interplay 
between macroscopic and microscopic aspects of interaction in hot, high-spin 
nuclear matter. In this sense, fission mechanism of heavy nuclei has been 
the subject of investigation for many years. The way how collective nature 
of the nuclear fission is modified, according to the incident projectile 
type, the transferred energy, and the momentum as well, is a question of 
interest to both experimentalists and theorists. A considerable amount of 
measurements have been performed for nucleon-induced fission on different 
heavy nuclei \cite{pro01}. Notwithstanding, a survey on the 
literature has displayed a considerable lack of experimental data for 
heavy-ion induced fission. There are some works on fission induced by 
carbon \cite{muk07,sag03} and $\alpha$-particles \cite{ing00} projectiles.  
For for lithium projectiles, a comprehensive investigation on fission 
of $^{208}$Pb nucleus induced by $^{6}$Li-ion beam, in the energy range of 
74.8-94.4 MeV, has been performed by Vigdor {\it et al.} \cite{vig82}, 
where they quantitatively studied the decay 
properties of heavy nuclei, at high angular momentum and high excitation 
energy. It is of particular interest to investigate the dependence of the 
fission cross-section  with the angular momentum transferred in the entrance 
channel when the fission is induced by nucleon- or heavy-ion projectiles. 
According to the classical rotating-liquid-drop model (RLDM) \cite{coh74} 
and various  other theoretical presentations \cite{sie86,wil73,cap97,mus75}, 
the fission barrier height should  decrease monotonically, and eventually 
vanishes, as the transferred angular momentum increases. Furthermore, 
investigation on fragments production due to fission induced by accelerated 
heavy ion can be of interest as possible application for the production of 
radioactive beams. High-energy fission induced by heavy nuclei could also 
be of interest for stimulated accelerator-driven systems (ADS's) for 
transmutation of nuclear waste, energy production, or other purposes where 
it is important to have knowledge of such nuclear reactions.

In this paper we present new results of a re-analysis of the data on fission 
reaction of lead target induced by $^{7}$Li ion beam at 245 MeV.  The 
previous analysis was presented in ref. \cite{dem05} with some incorrect  
cross-sections due to the misuse of fission fragment registration efficiency. 
The cross sections for fission fragments between mass $60<A<150$ were then 
re-evaluated using charge and mass distributions and are present here in 
table 1. This mass region is assumed to have most of the fission fragments 
from binary fission since, for heavy nuclei, the fission is characterized 
by a broad mass distribution and it is the most probable decay channel 
in the energy range considered here.
The new detection efficiencies considered in the present paper allowed us to 
calculate cross-section for 9 new fission fragments in the mass range of 
A=61-69 a.m.u. They are $^{61}$Cu, $^{62}$Zn, $^{65}$Ni, $^{65}$Zn, $^{65}$Ga, 
$^{66}$Ni, $^{66}$Ga, $^{67}$Ga and $^{69}$Ge.  
Cross sections for some elements, $^{69m}$Zn, $^{77}$Kr, $^{85m}$Y, $^{87}$Kr, 
$^{87(m+g)}$Zr, $^{89m}$Nb, $^{92m}$Nb, $^{95m}$Tc, $^{95}$Ru, $^{96}$Nb, $^{101}$Pd, 
$^{105}$Ru, $^{105(m+g)}$Ag, $^{105}$Cd, $^{111m}$Pd, $^{111}$In, $^{111}$Sn, 
$^{117g}$In, $^{121}$Xe, $^{126(m+g)}$Sb, $^{127g}$Sn, $^{127}$Cs, $^{133}$La, 
$^{136}$Cs, $^{137m}$Ce, $^{141}$Ce, $^{147}$Eu, $^{148g}$Pm, $^{148m}$Pm and
$^{148}$Eu, presented in the previous paper were estimated to be just up to 
upper limits of their production this time, i.e., their photopeak yields 
were comparable to the background and they were dropped out from the present 
analysis. Moreover, cross-sections for 29 elements present in the previous 
and present analysis, ${^67}$Cu, $^{71m}$Zn, $^{72}$Zn, $^{72}$Ga, $^{73}$Ga, 
$^{75}$Se, $^{75}$Br, $^{76}$As, $^{77(m+g)}$Br, $^{78}$Ge, $^{78}$As, $^{83}$Rb, 
$^{84}$Br, $^{84(m+g)}$Rb, $^{85g}$Sr, $^{87m}$Y, $^{89}$Rb, $^{92}$Sr, $^{92}$Y, 
$^{94}$Y, $^{95}$Zr, $^{95(m+g)}$Nb, $^{95g}$Tc, $^{96(m+g)}$Tc, $^{97}$Ze, 
$^{121}$I, $^{124(m+g)}$Sb, $^{133m}$Ba, $^{140}$Ba, were re-calculated.  
Some of the cross-sections were changed by the factor 6-7, some – by 3-4.
Cross-sections for some light nuclides fragments were also presented 
in the previous publication but their re-analysis will be considered 
in a further paper.  Moreover, based on the dynamic model \cite{wil73}, 
it was possible to extract the  angular momentum imparted to the fissioning 
nucleus at the interaction initial stage.  The results were compared to the 
fission data of other systems also induced by heavy nuclei.  In section-2 we 
describe the experimental procedure. In section-3 we present data analysis 
and the results on cross sections, while in section-4 we give some details 
on kinematic analysis of the reaction. Section-5 is devoted to the conclusion.

\section{Experimental Procedure}

The data here analysed was obtained by a measurement carried out by 
bombarding a natural lead target 
with an accelerated $^{7}$Li-ion beam of energy 35 MeV/u from the U-400M 
cyclotron in the Joint Institute for Nuclear Research (JINR), Dubna, Russia. 
The target is composed of an assembling of seven foils, all together, of pure 
lead, having a natural isotopic composition ($^{nat}$Pb: 1.48\% $^{204}$Pb, 
23.6\% $^{206}$Pb, 22.6\% $^{207}$Pb, and 52.3\% $^{208}$Pb), and a total 
thickness of 12 $\mu$m. The irradiation time was 40 min at 
ion beam intensity of about 10$^{10}$ nuclei per second. Two aluminum foils, 
20 $\mu$m thick each, located at forward and backward directions relative to
the beam, providing a solid angle of about 2$\pi$,
were used to collect the recoiled fragments from the 
fission on both sides of the target. 
The thickness of the catcher Aluminum foil was chosen to be larger than the 
longest recoil range of the fragments. The activity of the fission fragments 
were analysed in off-line mode by using the induced-activation method. The 
ranges and ratio of the emitted radioactive reaction products at 
forward and backward directions were also recorded. The spectra for the 
$\gamma$-rays emitted in the decays of radioactive fission products were 
measured from 10 min after the completion of the irradiation and lasted five 
months.  HpGe detector with energy resolution of 0.23\% at 1332 keV was used. 
The energy-dependent detection efficiency of the HpGe detector was measured 
with standard calibration sources of $^{22}$Na, $^{54}$Mn, $^{57;60}$Co, 
$^{137}$Cs,  $^{154}$Eu, $^{152}$Eu, and $^{133}$Ba. The half-lives of identified 
isotopes were within the range between 15min and 1yr.

The error in determining cross sections for each fission fragments depended 
on the following factors: the statistical significance of experimental 
results ($\leq$ 2-3\%), the accuracy in measuring the target thickness and 
the accuracy of tabular data on nuclear constants ($\leq$ 3\%), and the errors 
in determining the energy dependence of the detector efficiency ($\leq$ 10\%). 
Moreover, nuclear properties, such as nuclear transition energies, 
intensities, and half-lives, used to identify the observed isotopes, were 
taken from ref. \cite{fir96}.

The reaction fragment production cross-sections, in the absence of a parent 
isotope, that may give a  contribution in measured cross-section 
via $\beta^{\pm}$- decays, are usually considered as an independent 
yield (I) and it is determined by the following equation:

\begin{eqnarray}
\hspace{-0.2cm}\sigma=\frac{\Delta{N}\;\lambda}{N_{b}\,N_{n}\,k\,\epsilon\,\eta\,(1-\exp{(-\lambda
t_{1})})\exp{(-\lambda t_{2})}(1-\exp{(-\lambda t_{3})})},\label{g1}
\end{eqnarray}

\noindent
where $\sigma$ is the cross-section of the reaction fragment production (mb); 
$\Delta{N}$ is the yield of the photo-peak; $N_{b}$ is the number of 
particles in the beam (min$^{-1}$); $N_{n}$ is the number of target nuclei 
(1/cm$^{2}$); $t_{1}$ is the irradiation time; $t_{2}$ is the time of exposure 
between the end of the irradiation and the beginning of the measurement; 
$t_{3}$ is the time measurement; $\lambda$ is the decay constant (min$^{-1}$); 
$\eta$ is the intensity of $\gamma$-transitions; $k$ is the total coefficient 
of $\gamma$-ray absorption in target and detector materials, and $\epsilon$ 
is the $\gamma$-ray-detection efficiency.

In the case where the yield of a given isotope includes a contribution from 
the $\beta^{\pm}$-decay of neighboring unstable isobars, the cross section 
calculation becomes more complicated \cite{bab98}. If the formation 
probability for the parent isotope is known from experimental data, or if it 
can be estimated on the basis of other sources, hence the independent yields 
of daughter nuclei can be calculated by the relation:

\begin{eqnarray}
\sigma_{B}=&&\frac{\lambda_{B}}{(1-\exp{(-\lambda_{B}t_{1})})\,\exp{(-\lambda_{B}t_{2})}
(\,1-\exp{(-\lambda_{B} t_{3})})}\times\nonumber\\
&&\hspace*{-1.5cm}\left.\Biggl[\frac{(\Delta{N})_{AB}}{N_{\gamma}\,N_{n}\,k\,\epsilon\,\eta}-\sigma_{A}\,f_{AB}\,
\frac{\lambda_{A}\,\lambda_{B}}{\lambda_{B}-\lambda_{A}}
\Biggl(\frac{(1-\exp{(-\lambda_{A} t_{1})})\,\exp{(-\lambda_{A} t_{2})}\,(1-\exp{(-\lambda_{A} t_{3})})}\,
{\lambda^{2}_{A}}\right.\nonumber\\
&&\hspace*{-5.0cm}\left.\qquad\qquad\qquad\qquad\qquad\qquad\quad\qquad\qquad\qquad-\frac{(1-\exp{(-\lambda_{B} t_{1})})\,\exp{(-\lambda_{B}
t_{2})}\,(1-\exp{(-\lambda_{B} t_{3})})}\,{\lambda^{2}_{B}}\Biggr)\right.\Biggr],
\end{eqnarray}

\noindent 
where the subscripts $A$ and $B$ label variables refer to parent and daughter 
nucleus, respectively; the coefficient $f_{AB}$ specifies the 
fraction of $A$ nuclei decaying to a $B$ nucleus ($f_{AB}$ =1, when the 
contribution from the $\beta$-decay corresponds 100\%); and $(\Delta{N})_{AB}$ 
is the total photo-peak yield associated with the decays of the daughter and 
parent isotopes. The effect of the precursor can be negligible in some 
limiting cases:  where the half-life of the parent nucleus is very long, 
or in the case where its contribution is very small. In the case when parent 
and daughter isotopes could not be experimentally separated,  the calculated 
cross-sections are classified as cumulative (C). It should be mentioned 
that the induced-activity method used here imposes some
restrictions on the registration of the reaction products. The most severe
of them is that it is not possible to measure  stable and short-lived isotopes.
We overcome this limitation by considering the charge and mass distribution
analysis as explained in the following section.

\section{Experimental Data Analysis}

The experimental cross-sections of fission fragment production in the mass 
range of 60 $\leq A \leq$ 150 amu are presented in table 1 and fig. 1. The  
experimental points in the figure are the cross sections for each isobar chain 
of the observed fragments obtained by the induced-activity method. 
The solid black line is fission cross section for each mass number of fission 
products.  To obtain this fission mass distribution, it is necessary to 
estimate the cross-sections for the isotopes not measurable by the 
induced-activity method. In the present work, therefore, we obtain these 
cross-sections by analysing the charge and mass distributions.

The charge distribution is assumed to have the following Gaussian function 
\cite{bab98}:

\begin{eqnarray}
\sigma_{A,Z}=\frac{\sigma_A}{(C\pi)^{1/2}}\exp\left({-\frac{(Z-Z_p)^2}{C}}\right),
\end{eqnarray}

\noindent 
where $\sigma_{A,Z}$ is the independent cross-section for a given nuclide 
production with  atomic charge $Z$ and a mass number $A$; $\sigma_A$ is the 
isobaric cross-section of the mass chain $A$, $Z_p$ denotes the most 
probable charge for a given  isobar, and $C$ denotes the width parameter of 
the charge distribution.

This kind of analysis for charge distribution has been reported earlier for 
fission experiments induced by different projectile at low energies on heavy 
targets \cite{kud98,bra77,wah62}. What was observed in these works is that the 
charge distribution widths are  practically independent on the excitation 
energy, projectile and target nuclear properties. Therefore, we assumed that 
the width  parameter of the Gaussian is constant for different mass chains 
($C=1$). By fitting the distributions, using least-squares method, we obtained 
the most probable charge $Z_p$ and $\sigma_A$ parameters for a given isobar 
chain.  At first, only independent (I) yields are used. Hence, during 
successive approximation procedures, the estimation of the independent 
component of cumulative cross-section is extracted.  In table 2 we listed the 
most probable charges ($Z_p$) evaluated and their corresponding cross-sections. 
The values obtained here are in good agreement with data obtained for the 
fission of Bi target induced by $^{12}$C projectile \cite{bra77}.

Figures 2-4 shows the calculated fractional cross-sections 
(cross-section of fragment production to total cross-section for given mass 
number), as well as their Gaussian charge distribution, for different isobaric 
chains, as a function of the difference ($Z_p-Z$). It can be proposed that 
charge distributions of fission fragments are mainly determined by the  
properties of forming nuclei.

To obtain the total fission cross section we analysed the mass distribution 
of fission fragments as a function of A. This mass distribution is assumed to 
be symmetric and given by a Gaussian function defined as \cite{dui99}:

\begin{eqnarray}
\sigma_f=\lambda_A\exp\left({-\frac{(A-M_A)^2}{\Gamma_A^2}}\right), 
\label{roldao}                                                                  
\end{eqnarray}

\noindent
the parameters $\lambda_A$, $M_A$ and $\Gamma_A$ 
 stand for the height, mean mass $M_A$ and width, respectively. 
The result of the sum over all mass numbers of the fission fragments allows  
the estimation of the total fission cross-section.  To avoid double counting
due to the two fission fragments in each event, the sum of all cross-sections,
was multiplied  by a factor 0.5. The mass yield distribution obtained by the 
present fitting procedure is shown in fig. 1 (solid black curve).  The values 
of the fit parameters,  together with the obtained fission cross-section,  
are tabulated in table 3. The present mass distribution shows a maximum 
around mass $A=103$. Therefore a mass $A=206$ is expected for the 
fissioning  nucleus in contrast with mass $A=215$ of the compound nucleus. 
This indicates that about 9 nucleons (presumably neutrons), in average, were 
emitted  during the fission process. This is reasonable, since a compound 
nuclear reaction of a charged particle and a heavy nucleus leads to an 
excitation energy well above the fission threshold (top of fission barrier).

\section{Kinematics analysis}

The kinematic analysis of the  measured recoil  nuclei were also performed
in the frame of the two step vector model \cite{win80}, described in detail in 
\cite{dem05}.  The kinetic energy $T$ associated to each fragment can be 
determined from the recoil range values in the target, and with the help of 
the  Northcliffe-Schilling tables \cite{nor70}.  To transform the ranges into  
kinetic  energy of fission fragments we used the relation: 

\begin{eqnarray}
R =KT^{N/2},  
\end{eqnarray}

\noindent where the parameters $K$ and $N$ are obtained, for a given $Z$ 
and $A$, by fitting the range dependence on energy of accelerated ions. 
By its turn, to obtain the recoil range in the target material, $R=2W(F+B)$, 
as well as the forward-backward ($F/B$) ratio of product emission, 
where here $W$ is the thickness of the target in mg/cm$^2$, we have taken the 
relative amount of radioactive products in the forward ($F$) and backward 
($B$) at the catcher foils. These amounts are given by:

\begin{eqnarray}
F=\frac{S_F}{S_F+S_B+S_T},   \qquad B=\frac{S_B}{S_F+S_B+S_T},
\end{eqnarray}

\noindent where $S_F$ , $S_B$, and $S_T$ are the photo-peak yields 
associated with the  radioactive products from the catcher foils 
(forward and backward)  and in the target.  

Also, from the kinetic energy, we can estimate the mean excitation energy 
and consequently the corresponding relative parallel momentum of each pair 
of fragments, $p_{\parallel}/p_{CN}$,  where $p_{CN}$ is the momentum of a 
hypothetical compound nucleus formed in a complete fusion. They are related 
 by the following expression \cite{lag79}:

\begin{eqnarray}
\frac{E^{*}}{E_{CN}}=0.8\frac{p_{\parallel}}{p_{CN}}     \label{7}         
\end{eqnarray}

These kinematic parameters for some fragments obtained in present work 
are given in table 4.  The total kinetic energy released in each fission 
process can be determined as a sum of the energies of two presumed fragments. 
For the present fission process the total kinetic energy is estimated
to be 141$\pm$13 MeV, on average, which is in fairly good agreement with 
142.7 MeV estimated from the statistical approximation \cite{vio85}.

The energy regime of fission reaction induced by heavy particles is an 
important parameter to discuss the mechanism of the process. Reactions where 
incident particle velocity is higher than Fermi velocity of the nucleons in 
the compound nucleus, the individual nucleon-nucleon interaction dominates 
over the nucleus-nucleus interaction in the entrance channel, affecting the 
statistical description of the reaction.  Saint Laurent {\it et al} 
\cite{sai82} proposed that an important kinematic parameter to describe the 
change of the energy regime is the linear momentum imparted to the target 
nucleus. They have plotted the average parallel momentum transferred as a 
function of the projectile incident energy, both quantity divided by the mass 
number of the projectile, for available data on fission reactions induced 
by protons, alphas, deuterium on Thorium and Uranium as well as heavy 
incident particles such as  $^{16}$O and $^{20}$Ne on Uranium, and compared 
with the full momentum transferred, see their fig. 2 in ref. \cite{sai82}. 
Full momentum transferred would correspond to the momentum for the compound
nucleus formation and smaller transferred momentum may indicate the 
system lacks of complete fusion process. The authors in ref. \cite{sai82} 
have then proposed that the energy regime for fission induced by heavy 
particles could be divided in three energy regions: (i) low-energy range, 
$E/A\leq$ 10 MeV/u and $p_{\parallel}/p_{CN}$ $\sim$1, where complete fusion 
is the dominate process; (ii) intermediate energy range,  10 MeV/u 
$\leq E/A \leq$ 70 MeV/u, where $p_{\parallel}/p_{CN}$ decreases progressively 
down to 0.5, i.e.,  about half of the incident beam momentum is transferred 
to the target. At such energies, there is an indication that transferred 
momentum is proportional  to the mass of the projectile; (iii) high energy 
range,  70 MeV/u $\leq$ E 1000 MeV/u, where there is a drastic decrease of the 
momentum transferred to the target as compared with full momentum transferred. 

In our case, for  $E/A=35 (MeV/u)$, we are in the regime region ii).
The average value of the linear momentum transferred divided 
by full momentum transferred to the target obtained 
in the present study is $p_{\parallel}/p_{CN}=0.46 \pm 0.09$.
Our data support the suggestion that for the energy regime between  
10 MeV/u and  70 MeV/u,  the $p_{\parallel}/p_{CN}$ decreases down to 0.5, 
but remains approximately independent of the target mass and of the 
projectile identity. Thus, our data  can confirm that a complete fusion 
is not the dominant process at the energy regarded.
In this intermediate energy region, the complete fusion process at 
low limits are replaced partially by other processes, as the energy increases. 
We can say that only approximately 40\% of the fission channel at this 
energy can result from complete fusion. The remain momentum transferred
could be connected to incomplete fusion, or a significant pre-equilibrium 
contribution of the  emission of nucleons or light nuclei.

The mean excitation energies for the present reaction was estimated to be 
$E^{*}$ = 90$\pm$18 MeV,  correspond to nuclear temperature  
$T$ = 1.9$\pm$0.2 MeV.  Excitation energy and  temperature are related by the 
expression a$T^{2}$+4$T$ = $E^{*}$  \cite{bla54}, where $a=A/8$ is the 
level-density parameter and $A$ is the mass of the compound nucleus.
Our value for the mean excitation energy is similar to the one
obtained for the fission reaction of $^{6}$Li+$^{208}$Pb system at 
94.4 MeV \cite{vig82},  $E^{*}$= 100 MeV ($T= 2.1$ MeV). We can conclude
that, in our case, where the incident energy for $^7$Li beam is higher,
the contribution of the fusion process is much reduced
as the proposition asserted by authors in \cite{sai82}.

The fission cross-section in the present work was found to be 605$\pm$91 mb. 
The ratio between the fission cross-section $\sigma_f$ and the fission 
cross-section for the proton-induced fission at the same excitation energy 
\cite{pro01} is a factor 5 times higher. Such fission cross-section increment 
can be associated with  a higher angular momentum transferred by the  
accelerated heavier ion during the fission process.  This can be related to 
a higher rotation energy impinged by heavier projectile to the compound 
nucleus. As it is well known from different macroscopic and microscopic models 
\cite{coh74,sie86,wil73,cap97,mus75}, the rotational energy  of hot nuclei 
strongly affects the fission probability. According to the dynamic 
model \cite{wil73}, the initial relative kinetic energy and angular momentum  
of the projectile, in a  heavy-ion induced fission reaction, is  converted 
into the intrinsic excitation energy and the spin of the fused compound 
system. The equilibrium is determined by the balance of the macroscopic 
surface energy, Coulomb energy, and the rotational energy. The concept of 
dynamical force equilibrium in the entrance reaction channel serves as a 
criteria to the determination of the average and maximum angular momentum 
transferred for different colliding systems. According to the model,  
transferred linear momentum should be a decreasing function of the impact 
parameter $b$. Furthermore, the value of a maximum angular momentum 
$\ell_{max}$ corresponds to a peripheral collision is defined by:

\begin{eqnarray}
\ell_{max}=R\sqrt{\frac{2\mu(E_{c.m.}-V_{CB})}{\hbar^{2}}}.     
\end{eqnarray}

Here, $R$ is the maximum distance between two nuclei at which the collision 
leads to a reaction, $\mu$ is the reduced mass and $V_{CB}$ is the Coulomb 
energy of the system at distance $R$; $E_{c.m.}$ is at the center-of-mass 
bombarding energy.

For the present study for the $^7$Li+Pb system at $E_{c.m.}=237$ MeV, the 
maximum angular momentum is $\ell_{max} \sim 90\, \hbar$. The average angular 
momentum imparted into the fissile system at energies above the barrier can 
also be derived using the following equation \cite{cap97}:

\begin{eqnarray}
< \ell >=\frac{2}{3}\sqrt{\frac{2\mu R^2(E_{c.m.}-V_{CB})}{\hbar^2}}.  
\label{9}  
\end{eqnarray} 

In our case, the average angular momentum is estimated to be about
$< \ell >$  = 55-60 $\hbar$.  The interval is because the spin distribution 
is not a sharp cutoff. For comparison, the value of the average angular 
momentum obtained for the $^6$Li+$^{208}$Pb system at 94.4 MeV, corresponding
to a complete fusion cross section, is  30-35 $\hbar$.  Also, for a higher 
incident energy of the  projectile, as our case, the number of evaporated 
neutrons should increase, opening new fission channels. As a result, not 
only the fission cross section should be higher but we also observed a broad 
distribution of transferred linear momentum and excitation energy, as shown in 
table 4. For  $^6$Li+$^{208}$Pb system at 94.4 MeV the average angular momentum 
is lower,  and therefore it seems that the entire linear momentum is 
transferred to the compound nucleus.  The variety of fragments in the present 
work with high and low linear  momentum   is also an indication of possible 
incomplete-fusion mechanism, as expected for a system in such range of energy. 

As known, the presence of angular momentum lowers the fission-barrier heights
\cite{coh74,sie86,wil73,cap97,mus75}
 and it can be used to analyze the energy dependence of the excited 
nucleus fissibility \cite{mas93}. In our case, fission is expected to play a 
significant role for angular momentum above 50 $\hbar$, when the 
fission barrier has dropped to about half of the previous value, at zero 
angular momentum.  The fission-barrier dependence on the angular momentum 
can be neglected in the case of nucleon-nucleus interaction at intermediate 
energies, where only a small angular momentum is imparted to the compound
nucleus by the incident nucleon. On the another hand, it becomes relevant for 
higher values for $\ell$, in heavy-ion-induced reactions. It provides a 
careful and cogent explanation for the higher fission cross-section in the 
present investigation as compared with fission induced by protons or neutrons.

\section{Conclusion}

New analysis of the data regarding the fission process of $^{7}$Li+$^{nat}$Pb 
at intermediate energies 245 MeV is presented. The fission cross-section, 
derived from the analysis of the obtained charge and mass distributions 
was equal to 605$\pm$91 mb.  This value exceeds the cross section obtained 
for fission induced by protons at approximately the same energy range. 
This can be explained by the effect of the transferred angular momentum,
which is much higher for heavy ion-induced reaction.

The measurement of the recoil fission fragments,  and by considering their 
kinematic properties, in the frames of two step models, allowed the 
calculation of some important reaction parameters such as relative transferred 
momentum. The average relative value of the transferred momentum  
$p_{\parallel}/p_{CN}$  (where $p_{CN}$ is the total momentum of the hypothetical 
complete nucleus), obtained in the present analysis is 0.46$\pm$0.09. This 
value contain the information on the initial reaction mechanism, and 
indicates that, in the  energy regime investigated, the fission process 
does not  proceed solely via the compound nucleus formation. Other mechanisms 
may take participation on the first step reaction, and only ~40\% of all 
fission events are resultant of compound nucleus fission.

Regarding the dynamic model-based calculation, the fissile system is assumed 
to be formed at an intermediate energy with high angular momentum. This might
be the explanation for the  high value of fission cross-section for 
heavy-ion-induced fission.

G. Karapetyan would like to thank Funda\c c\~ao de Amparo \`a Pesquisa do 
Estado de S\~ao Paulo (FAPESP) 2011/00314-0 for the financial support
and International Centre for Theoretical Physics (ICTP) under the Associate 
Grant Scheme.

\newpage
\begin{table}
\caption{Cross-sections of fission fragments. $C$ and $I$ stands for 
independent and cumulative cross sections as explained in the text.}
\begin{center}
\begin{tabular}{||c|c|c||c|c|c||} 
\hline
Element &Type&$\sigma$, mb&Element&Type&$\sigma$, mb\\ \hline
$^{61}$Cu& C &$\leq$0.11&$^{99}$Mo& C &10.7 $\pm$1.1\\ \hline
$^{62}$Zn& C &$\leq$0.04&$^{99m}$Tc& I & 0.26$\pm$0.03\\ \hline
$^{65}$Ni& I &$\leq$0.20&$^{101m}$Rh& I &0.78 $\pm$0.08\\ \hline
$^{65}$Zn& I &0.40$\pm$0.08&$^{102m}$Rh& I &0.42$\pm$0.03\\ \hline
$^{65}$Ga& C &$\leq$0.02&$^{103}$Ru& C &15.7 $\pm$1.5 \\ \hline
$^{66}$Ni& I &0.11$\pm$0.02&$^{105(m+g)}$Rh& I  & 13.7 $\pm$1.4\\ \hline
$^{66}$Ga& I &$\leq$0.035&$^{106}$Ru& C &11.1$\pm$1.1\\ \hline
$^{67}$Cu& C &0.72$\pm$0.07&$^{106m}$Rh& C &7.60$\pm$0.91\\ \hline
$^{67}$Ga& C &0.026$\pm$0.003&$^{110m}$Ag& I & 2.70$\pm$0.40\\ \hline
$^{69}$Ge& C&$\leq$0.45&$^{111(m+g})$Ag& C &10.2$\pm$1.0\\ \hline
$^{71m}$Zn& I &0.96$\pm$0.14&$^{111}$Cd& I &$\leq$0.63\\ \hline
$^{72}$Zn& I &$\leq$0.02&$^{112}$Pd& C &3.10 $\pm$0.31\\ \hline
$^{72}$Ga& I &0.25$\pm$0.02&$^{112}$Ag& I &2.10$\pm$0.25\\ \hline
$^{73}$Ga& I &0.63$\pm$0.08&$^{113(m+g)}$Ag& C  & 2.45$\pm$0.37\\ \hline
$^{74}$As& I &0.37$\pm$0.04&$^{113(m+g)}$Sn& C  & 3.34$\pm$0.50\\ \hline
$^{75}$Se& I &0.19$\pm$0.01&$^{115}$Cd& C &1.25$\pm$0.13\\ \hline
$^{75}$Br& C &$\leq$0.017&$^{117g}$Cd& C &1.26 $\pm$0.13\\ \hline
$^{76}$As& I &2.77$\pm$0.28&$^{117m}$Cd& C &1.31$\pm$0.16\\ \hline
$^{77(m+g)}$Ge& C &2.00$\pm$0.20&$^{117m}$Sn& I &2.83$\pm$0.28\\ \hline
$^{77(m+g)}$Br& I &$\leq$0.02&$^{118(m+g)}$Sb& I &1.90$\pm$0.20\\ \hline
$^{78}$Ge& I &2.40$\pm$0.24&$^{120m}$Sb& I &3.80$\pm$0.40\\ \hline
$^{78}$As& C &$\leq$1.98&$^{121g}$Te& I & 4.60$\pm$0.69\\ \hline
$^{81m}$Se& C &$\leq$0.77&$^{121m}$Te& I &1.95$\pm$0.29\\ \hline
$^{82(m+g})$Br& I &1.67$\pm$0.17&$^{121}$I& I &$\leq$0.45\\ \hline
$^{83}$Rb& C &1.85$\pm$0.19&$^{122(m+g)}$Sb& I &2.74$\pm$0.30\\ \hline
$^{84}$Br& C &4.80$\pm$0.50&$^{123m}$Te& I &4.73$\pm$0.50\\ \hline
$^{84(m+g)}$Rb& I &1.14$\pm$0.11&$^{123}$I& I &0.94$\pm$0.10\\ \hline
$^{85g}$Sr& I & 1.80 $\pm$0.18&$^{123}$Xe& C &$\leq$0.04\\ \hline
$^{85m}$Sr& I &$\leq$0.34&$^{124(m+g)}$Sb& I &2.60$\pm$0.26\\ \hline
$^{85g}$Y& C & 0.35 $\pm$0.04&$^{124}$I& I &1.25$\pm$0.13\\ \hline
$^{86(m+g)}$Rb& I &7.0$\pm$1.1&$^{125}$Sb& C &0.86$\pm$0.09\\ \hline
$^{87g}$Y& I & 0.68$\pm$0.07&$^{126}$I& I &1.82$\pm$0.19\\ \hline
$^{87m}$Y& C & 0.50$\pm$0.06&$^{127}$Sb& C &$\leq$0.06\\ \hline
$^{88}$Y& C & 1.12$\pm$0.35&$^{127(m+g)}$Xe& I &0.29$\pm$0.03\\ \hline
$^{89}$Rb& C & 2.20$\pm$0.22&$^{128}$Sb& C &0.38$\pm$0.04\\ \hline
$^{89(m+g)}$Zr& C &0.48$\pm$0.05&$^{129}$Sb& C &$\leq$0.55\\ \hline
$^{90m}$Y& I & 3.80$\pm$0.40&$^{129m}$Te& I &1.62$\pm$0.17\\ \hline
$^{91}$Sr& C &4.94 $\pm$0.50&$^{129g}$Ba& C & $\leq$0.73\\ \hline
$^{91m}$Y& I &$\leq$0.32&$^{129m}$Ba& C &$\leq$0.19\\ \hline
$^{92}$Sr& C & 5.80 $\pm$0.58&$^{130(m+g)}$I& I &0.29$\pm$0.03\\ \hline
$^{92}$Y& I & 2.00 $\pm$0.20&$^{131m}$Te& C &$\leq$0.50\\ \hline
$^{93}$Y      & C & 13.0$\pm$1.3&$^{132}$Te& C &$\leq$0.02\\ \hline
$^{94}$Y      & C & 8.90$\pm$0.90&$^{132}$Cs& I &0.37$\pm$0.04\\ \hline
$^{95}$Zr     & C & 12.0 $\pm$1.2 &$^{133(m+g)}$I   & C & $\leq$1.00\\ \hline
$^{95(g+m)}$Nb & I & 9.8$\pm$1.0   & $^{133m}$Ba    & I & 0.30$\pm$0.06\\ \hline
$^{95g}$Tc    & I & $\leq$0.10    & $^{139}$Ba     & C & $\leq$0.45   \\ \hline
$^{96(g+m)}$Tc & I & $\leq$0.03    & $^{139(m+g)}$Ce & C & 1.17$\pm$0.12\\ \hline 
$^{97}$Zr     & C & 3.12$\pm$0.31 & $^{140}$Ba     & C & 0.33$\pm$0.03\\ \hline 
$^{97(g+m)}$Nb & I & 3.55$\pm$0.36 & $^{140}$La     & I & 0.18$\pm$0.02\\ \hline
$^{97}$Ru     & C & $\leq$0.20    & $^{141}$La     & I & $\leq$0.3    0\\ \hline
$^{98m}$Nb    & I & 1.50$\pm$0.16 & $^{147}$Gd     & C & $\leq$0.03  \\ 
\hline
\end{tabular}
\end{center}
\end{table}

\begin{table}
\caption{Values of the most probable charge $Z_{p}$ and their corresponding
cross sections.}
\begin{center}
\begin{tabular}{||c|c|c||c|c|c||}  \hline
$A$   & $Z_{p}$ & $\sigma_{A}$ & $A$ & $Z_{p}$ & $\sigma_{A}$\\ \hline 
76  & 33.0$\pm$ 0.7 &  4.0 $\pm$ 0.4 & 110 & 45.7 $\pm$ 0.7 & 30.0 $\pm$0.8\\
\hline
77  & 32.5$\pm$ 0.5 &  4.0 $\pm$ 0.3 & 111 & 46.3 $\pm$ 0.6 & 30.0 $\pm$1.9 \\
 \hline
78  & 32.4$\pm$ 0.3 &  5.4 $\pm$ 0.9 & 112 & 45.6 $\pm$ 0.5 & 27.0 $\pm$1.5 \\
 \hline
82  & 34.0$\pm$ 0.2 &  9.6 $\pm$ 0.8 & 113 & 48.6 $\pm$ 0.5 & 32.0 $\pm$0.9 \\
 \hline
83  & 35.9$\pm$ 0.2 & 11.0 $\pm$ 0.5 & 115 & 49.4 $\pm$ 0.7 & 24.0 $\pm$0.9 \\
 \hline
84  & 35.6$\pm$ 0.3 & 12.0 $\pm$ 0.6 & 117 & 49.2 $\pm$ 0.6 & 20.0 $\pm$1.3 \\
 \hline
85  & 36.7$\pm$ 0.5 & 14.0 $\pm$ 0.7 & 118 & 49.7 $\pm$ 0.5 & 18.8 $\pm$0.6 \\
 \hline
86  & 36.5$\pm$ 0.5 & 15.5 $\pm$ 0.5 & 120 & 50.1 $\pm$ 0.3 & 15.0 $\pm$0.9 \\
 \hline
89  & 37.3$\pm$ 0.5 & 21.0 $\pm$ 1.4 & 121 & 51.5 $\pm$ 0.6 & 15.0 $\pm$0.8 \\
 \hline
90  & 37.8$\pm$ 0.5 & 24.0 $\pm$ 0.5 & 121 & 51.5 $\pm$ 0.5 & 15.0 $\pm$0.8 \\
 \hline
91  & 37.0$\pm$ 0.3 & 24.0 $\pm$ 1.2 & 122 & 52.0 $\pm$ 0.3 & 12.0 $\pm$0.9 \\ 
\hline
92  & 37.6$\pm$ 0.4 & 28.0 $\pm$ 1.0 & 123 & 51.8 $\pm$ 0.4 &  9.0 $\pm$0.1 \\
 \hline
93  & 39.4$\pm$ 0.4 & 27.0 $\pm$ 1.2 & 124 & 51.8 $\pm$ 0.5 &  9.0 $\pm$0.3 \\
 \hline
94  & 39.5$\pm$ 0.5 & 30.0 $\pm$ 1.0 & 125 & 52.3 $\pm$ 0.4 &  8.0 $\pm$0.4 \\ 
\hline
95  & 40.4$\pm$ 0.3 & 25.0 $\pm$ 1.3 & 126 & 52.3 $\pm$ 0.3 &  6.00 $\pm$0.09\\
 \hline
96  & 40.1$\pm$ 0.2 & 23.0 $\pm$ 0.8 & 127 & 52.6 $\pm$ 0.4 &  6.0 $\pm$0.4\\ 
\hline
97  & 41.3$\pm$ 0.6 & 30.0 $\pm$ 1.1 & 128 & 52.7 $\pm$ 0.5 &  4.5 $\pm$0.3\\ 
\hline
99  & 41.4$\pm$ 0.4 & 34.0 $\pm$ 0.8 & 129 & 52.4 $\pm$ 0.3 &  5.0 $\pm$0.1\\ 
\hline
101 & 43.2$\pm$ 0.3 & 38.0 $\pm$ 1.1 & 130 & 54.4 $\pm$ 0.3 &  3.9 $\pm$0.1\\ 
\hline
102 & 43.1$\pm$ 0.5 & 37.0 $\pm$ 0.7 & 132 & 54.0 $\pm$ 0.4 &  2.00 $\pm$0.06\\
\hline
103 & 43.5$\pm$ 0.4 & 38.0 $\pm$ 0.7 & 140 & 56.2 $\pm$ 0.2 &  0.60 $\pm$0.04\\
 \hline
105 & 44.7$\pm$ 0.5 & 30.0 $\pm$ 0.6 & 141 & 56.2 $\pm$ 0.5 &  0.40 $\pm$0.04\\
 \hline
106 & 44.3$\pm$ 0.4 & 30.0 $\pm$ 1.1 & & &\\ 
\hline
\end{tabular}
\end{center}
\end{table}

\begin{table}
\caption{Values of the parameters obtained in the fitting 
procedure of the mass distribution and fission 
cross-section $\sigma_{f}$.}
\begin{center}
\begin{tabular}{||c|c||}  \hline
$\Gamma_{A}$&17.84$\pm$0.89\\ \hline
$\lambda_{A}$&38.3$\pm$2.3\\ \hline
$M_{A}$&103.0$\pm$0.2\\ \hline
$\sigma_{f}$ (mb)&605$\pm$91\\ \hline
\end{tabular}
\end{center}
\end{table}

\begin{table}
\caption{Kinematic parameters of fission fragments.}
\begin{center}
\begin{tabular}{||c|c|c|c|c|c||}  \hline
Nucleus & $F/B$ & 2$W(F + B)$, mg/cm$^2$ & $T$ (MeV) & $E^*$ (MeV) & p$_{\parallel}$/p$_{CN}$\\ 
\hline 
$^{72}$Zn     & 1.11$\pm$0.22 & 13.2$\pm$2.6   & 107$\pm$22 & 69$\pm$12 & 0.33$\pm$0.06\\ 
\hline
$^{77(m+g)}$Br & 1.16$\pm$0.14 & 13.4$\pm$1.6   & 100$\pm$12 & 82$\pm$12 & 0.42$\pm$0.06\\ 
\hline
$^{82(m+g)}$Br & 1.17$\pm$0.14 & 11.8$\pm$1.4   & 96$\pm$12  & 84$\pm$12 & 0.43$\pm$0.06\\ 
\hline
$^{83}$Rb     & 1.17$\pm$0.08 & 10.25$\pm$0.72 & 93$\pm$7   & 80$\pm$8 & 0.41$\pm$0.04\\ 
\hline
$^{84(m+g)}$Rb & 1.14$\pm$0.10 &  8.17$\pm$0.74 & 90$\pm$8   & 69$\pm$8 & 0.35$\pm$0.04\\ 
\hline
$^{87(m+g)}$Y  & 1.19$\pm$0.06 & 10.50$\pm$0.53 & 88$\pm$4   & 86$\pm$4 & 0.44$\pm$0.02\\ 
\hline
$^{91}$Sr     & 1.18$\pm$0.10 & 10.10$\pm$0.81 & 79$\pm$6   & 76$\pm$8 & 0.39$\pm$0.04\\ 
\hline
$^{95}$Zr     & 1.19$\pm$0.16 & 11.4$\pm$1.7   & 77$\pm$12  & 78$\pm$12 & 0.40$\pm$0.06\\ 
\hline
$^{95(g+m)}$Nb & 1.23$\pm$0.11 & 11.3$\pm$1.0   & 77$\pm$7   & 90$\pm$8 & 0.46$\pm$0.04\\ 
\hline
$^{96(g+m)}$Tc & 1.29$\pm$0.14 & 10.1$\pm$1.1   & 75$\pm$8   & 100$\pm$10 & 0.58$\pm$0.06\\ 
\hline
$^{99}$Mo     & 1.22$\pm$0.06 & 10.33$\pm$0.52 & 72$\pm$4   & 84$\pm$4 & 0.43$\pm$0.02\\ 
\hline
$^{105(m+g)}$Rh& 1.17$\pm$0.09 & 10.07$\pm$0.81 & 71$\pm$6   & 67$\pm$4 & 0.34$\pm$0.02\\ 
\hline
$^{111(m+g)}$Ag& 1.22$\pm$0.12 & 8.32$\pm$0.83  & 66$\pm$7   & 76$\pm$8 & 0.39$\pm$0.04\\ 
\hline
$^{115}$Cd    & 1.19$\pm$0.12 & 10.8$\pm$1.1 & 59$\pm$6     & 65$\pm$8 & 0.33$\pm$0.04\\ 
\hline
$^{117m}$Sn   & 1.43$\pm$0.17 & 7.69$\pm$0.92 & 54$\pm$6    & 119$\pm$16 & 0.61$\pm$0.08\\ 
\hline
$^{120m}$Sb   & 1.45$\pm$0.06 & 4.29$\pm$0.17 & 51$\pm$2    & 119$\pm$8 & 0.61$\pm$0.04\\
 \hline
$^{124}$I      & 1.65$\pm$0.18 & 7.95$\pm$0.87 & 45$\pm$5    & 150$\pm$15 & 0.77$\pm$0.08\\ 
\hline
\end{tabular}
\end{center}
\end{table}

\newpage
\begin{figure}
\epsfig{file=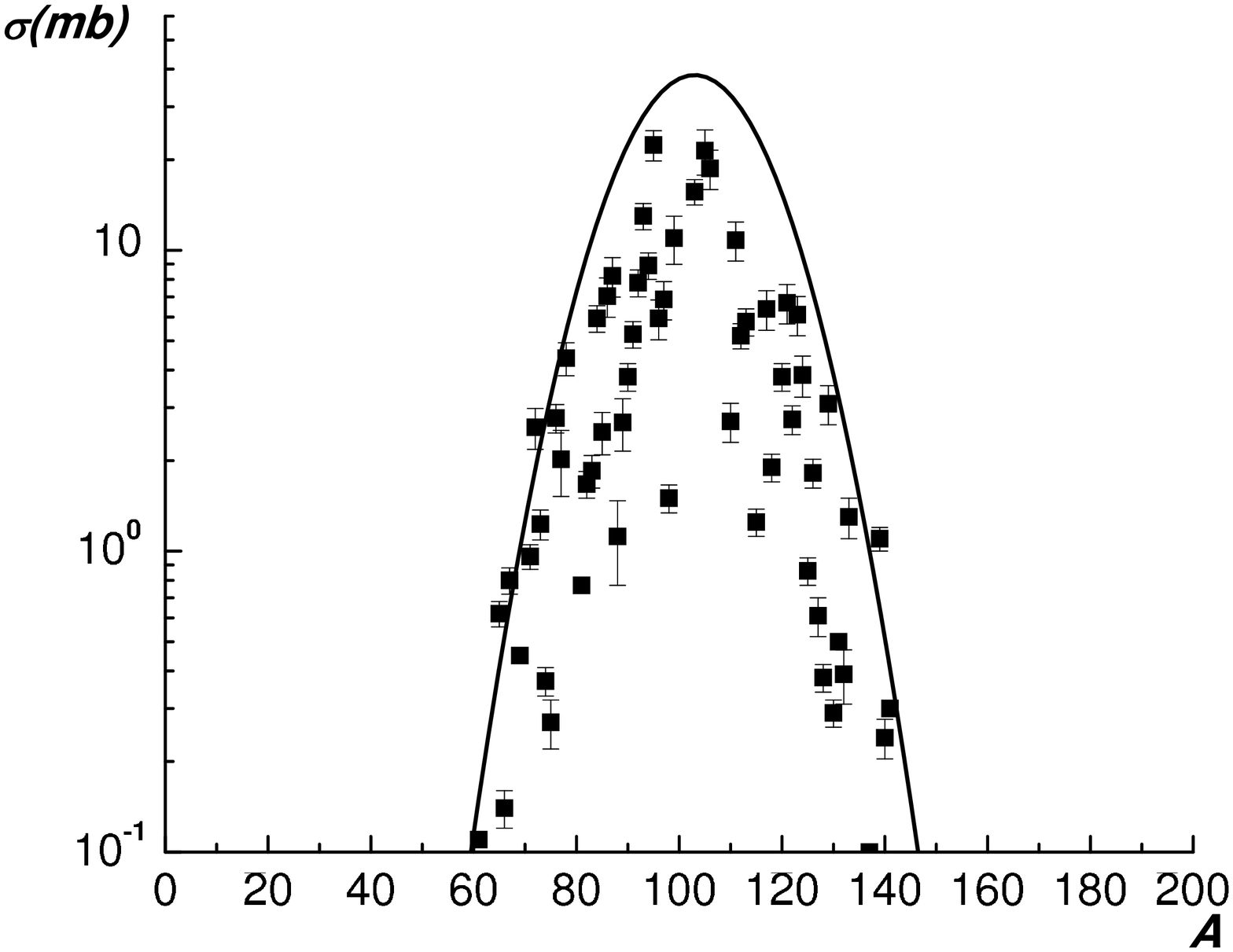,height=7cm,width=9cm}
\caption{Experimental cross sections  for the fission-products 
mass distribution of  245 MeV $^{7}$Li-induced fission on $^{nat}$Pb.  
The black continuous curve corresponds to the derived fission 
cross-section as a function of the mass of the fragments after
taking into account the estimation of the no-detected isotopes. }
\end{figure}

\begin{figure}
\epsfig{file=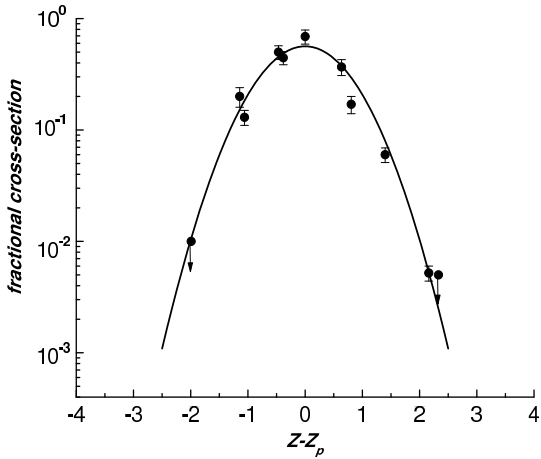,height=6cm,width=7cm}
\caption{ The charge distribution of fission products for the 245 MeV  
$^{7}$Li--induced fission on $^{nat}$Pb for isobaric chains in the mass region 
A=72-78.  The curves are the estimated Gaussian distributions.}
\end{figure}

\begin{figure}
\epsfig{file=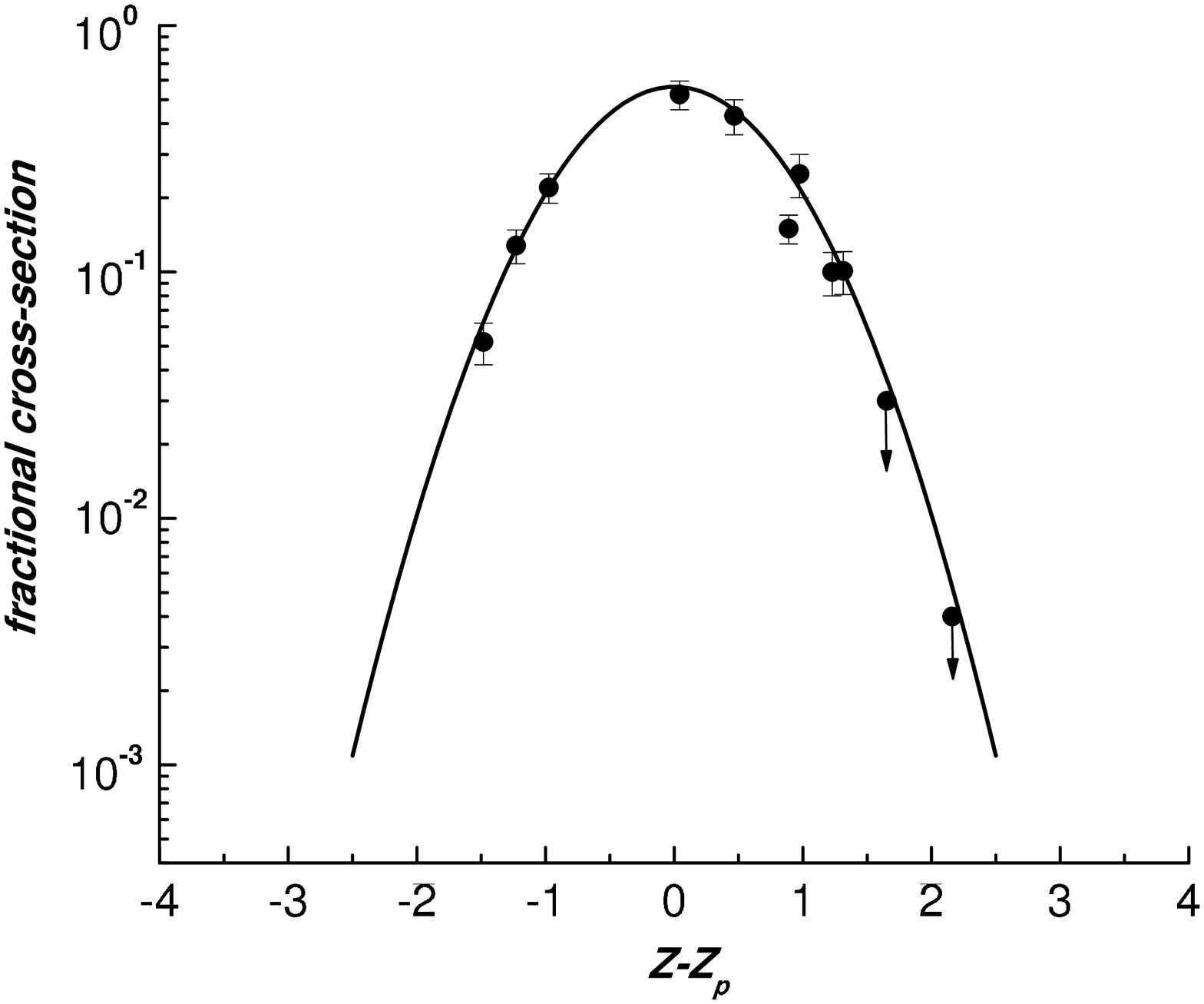,height=6cm,width=7cm}
\caption{ Idem for isobaric chains in the mass region A=115-118.}
\end{figure}

\begin{figure}
\epsfig{file=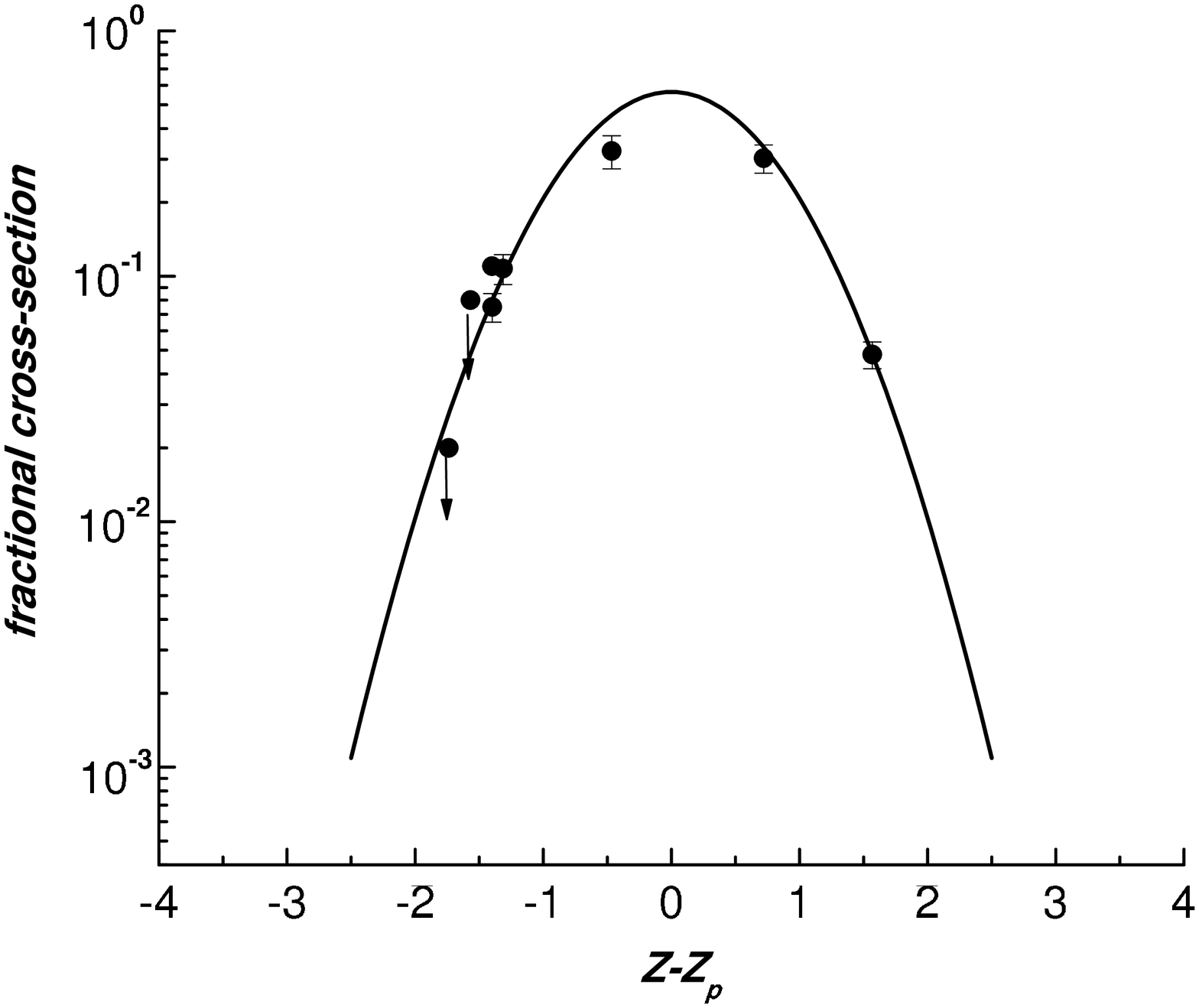,height=6cm,width=7cm}
\caption{ Idem for isobaric chains in the mass region A =125-130.}
\end{figure}

\end{document}